\documentclass[12pt,leqno]{amsart}
\newcommand{\R}{{\mathbb R}}
\newcommand{\N}{{\mathbb N}}

\def\C{{\mathbb C}} 

\def\wt{\widetilde }

\def\comp{{\rm comp}}

\def\cost{{\rm cost\, }} 

\def\e{\varepsilon } 
\def\epsilon{\varepsilon } 
 
\def\rho{\varrho } 
\def\l{{\lambda }} 
\def\phi{\varphi }
\def\a{\alpha }
\def\b{\beta }
\def\JC{J. Complexity } 
\def\({\biggl( } 
\def\){\biggr) }

\def\ran{{\rm random }} 
\def\quant{{\rm quant }} 
\def\coin{{\rm coin }} 
\def\query{{\rm query }} 

\def\ls{\left<}
\def\rs{\right>}
\def\lg{\left|}
\def\rg{\right|} 

\newcounter{Theorem} 

\theoremstyle{plain}
\newtheorem{theorem}[Theorem]{Theorem}

\theoremstyle{definition}

\newcount\refnum
\def\refx{\smallskip \global\advance\refnum by 1 {\the\refnum . \ }}

\begin{document}

\title[Quantum Complexity of Integration]{Quantum Complexity of Integration}

\author[Erich Novak]{Erich Novak}
\address{Mathematisches Institut, 
Universit\"at Jena, 
Ernst-Abbe-Platz 4, 
D-07740 Jena, Germany}
\email{novak@mathematik.uni-jena.de}

\date{August 19,  2000, revised: November 11, 2000. Appeared in J. Complexity {\bf
17} (2001), 2--16. Some minor mistakes are corrected in the present version.}

\begin{abstract} 
It is known that quantum computers yield a speed-up for certain discrete 
problems. Here we want to know whether quantum computers are useful for 
continuous problems. 
We study the computation of the integral of functions from the 
classical H\"older classes $F_d^{k,\a}$ on $[0,1]^d$ and define 
$\gamma$ by $\gamma = (k+\alpha)/d$. 
The known optimal orders for the complexity of
deterministic and (general) randomized methods are
$$ 
\comp( F_d^{k, \alpha}  , \e) \asymp \e^{-1/\gamma}
$$ 
and 
$$ 
\comp^{\rm random}  (F_d^{k, \a}  , \e) \asymp \e^{-2/(1+2\gamma)} .
$$ 
For a quantum computer we prove
$$ 
\comp^\quant_\query  (F_d^{k,\alpha}, \e) \asymp \e^{-1/(1+\gamma)}  
$$ 
and 
$$ 
\comp^\quant  (F_d^{k,\alpha}, \e) \le C \, \e^{-1/(1+\gamma)} \, 
(\log \e^{-1} )^{1/(1+\gamma)} . 
$$ 
For restricted Monte Carlo (only coin tossing instead of general 
random numbers) we prove
$$ 
\comp^\coin   (F_d^{k,\alpha}, \e) \le C \, \e^{-2/(1+ 2 \gamma)} \, 
(\log \e^{-1} )^{1/(1+ 2 \gamma)} . 
$$
To summarize the results one can say that 

\begin{itemize} 

\item
there is an exponential speed-up of quantum algorithms over 
deterministic (classical) algorithms, if $\gamma$ is small; 

\item
there is a (roughly) quadratic speed-up of quantum algorithms over 
randomized classical methods, if $\gamma$ is small.

\end{itemize} 

\end{abstract} 

\maketitle

\section{Introduction and Results} 

\subsection{Computation of the mean} 

Consider the following problem: 
Compute the mean 
$$
S_n (x)= \frac{1}{n} \sum_{i=1}^{n} x_i 
$$
of $n$ numbers $x_i$, where $|x_i | \le 1$, up to some error 
$0< \e <  1/2$.
The complexity of this problem depends on $n$ and $\e$ and 
in the real number model we obtain
\begin{equation}    \label{eq1}  
\comp  (n, \e) \approx n \cdot (1-\e) .
\end{equation}  
Here we consider the worst case setting, with the worst case cost and the 
worst case error. 
With randomized methods we can do much better, at least if $n$ is large 
compared to $\e^{-2}$. The cost is of the order
\begin{equation}    \label{eq2}  
\comp^\ran   (n, \e) \asymp  \min (n , \,  \e^{-2} ) .
\end{equation}  
Now the error of a method is a random variable and the requirement is 
that its expectation is bounded by $\e$. 
The statements \eqref{eq1} and \eqref{eq2} follow easily from well 
known upper and lower bounds. See, for example, Novak (1988). 

If we allow only random bits (restricted Monte Carlo methods, coin tossing) 
instead of arbitrary randomized methods then one gets the upper bound
\begin{equation}    \label{eq3}  
\comp^\coin   (n, \e) \le C \cdot \min (n , \,  \e^{-2} \log n   ) 
\end{equation}  
which follows easily from \eqref{eq2}. 

For the results (1--3), and for all classical algorithms, we use the real number 
model of computation, with unit cost for each arithmetic operation. 
In addition we allow for (2) the instruction 
``choose a random number $x \in [0,1]$''and for (3) the instruction 
``flip a coin'' or ``choose randomly $\{ 0,1 \}$'' and also the cost of 
these instructions is one. See Novak (1995) for more details. 

A further improvement is possible by a quantum computer. 
The upper bound $\e^{-1}$ for the query complexity, defined by the number of times the 
real valued oracle is accessed to solve the 
problem, is proved in 
Brassard, H\o yer, Mosca, Tapp (2000).
See Theorem 12 of this paper. 
Using also the lower bounds from 
Nayak, Wu (1998) one can see that the exact order of this query complexity is 
\begin{equation}    \label{eq4}  
\comp^\quant_\query (n, \e)  \asymp  \min (n , \,  \e^{-1} ) .
\end{equation}  
The bound \eqref{eq4} is very important for the present paper. 
Grover (1998) states the upper bound $\e^{-1}$ and says that it is optimal 
``up to polylogarithmic factors''. 

If we consider, for the quantum computer, the bit number model then we 
need a slightly larger cost. 
H\o yer (2000) proves the upper bound 
\begin{equation}    \label{eq5}  
\comp^\quant    (n, \e) \le C \cdot \min ( n \, , 
\e^{-1} \, (   \log n + \log  \log \e^{-1} ),   
\end{equation}  
see Section \ref{grover} for details on the model of computation. 
The term $n$ on the right side of (5) comes from the trivial classical algorithm. 
This term is $n \, (\log n + \log \e^{-1})$ in the classical bit number model. 
The output of a quantum algorithm is a random variable $A(x, \e)$, 
we always request that 
\begin{equation}    \label{eq6} 
|A(x, \e) - S_n(x)| \le \e \quad \hbox{ with probability at least $3/4$}. 
\end{equation} 
Of course we can run the algorithm several 
times and, taking the median from several measurements, we increase the 
probability of success. 

\medskip

\subsection{Computation of integrals} 

How can we apply these results to the problem of numerical integration? 
Let us first consider the computation of the integral
$$
I(f) = \int_{[0,1]^d} f(x) \, dx 
$$
for functions from the H\"older classes 
$$
F_d^\a = \{ f: [0,1]^d \to \R \mid \Vert f \Vert_\infty  \le 1, \ 
|f(x) - f(y) | \le \Vert x-y \Vert^\a \} , 
$$
$0 < \alpha \le 1$. 
Consider, for $d=1$, the midpoint rule 
$$
Q^1_\ell(f) = \frac{1}{\ell} \sum_{i=1}^\ell f \left( \frac{2i-1}{2\ell} \right) 
$$
and for $d>1$ the respective tensor product $Q^d_n$ that uses $n=\ell^d$ function values. 
By well known estimates for $d=1$ together with the technique of 
Haber (1970, p. 489) we get the estimate 
\begin{equation}              \label{eq7} 
e(Q^d_n, F_d^\alpha ) \le C \cdot d \cdot n^{-\alpha /d} 
\end{equation} 
for the worst case error of the product rule. To obtain $e(Q^d_n, F_d^\alpha) \approx 
\e$, we have to take 
\begin{equation}    \label{eq8}  
n(F_d^\alpha , \e) \approx \left( \frac{C \, d }{\e} \right)^{d/\alpha} 
\end{equation} 
function values. 
We can now use the results from above to obtain upper bounds for the complexity 
of numerical integration. 
Using the (trivial) result \eqref{eq1} we get 
a bound for the (worst case) complexity of integration,
\begin{equation}    \label{eq9}  
\comp( F_d^\alpha , \e) \le \left( \frac{C \, d }{\e} \right)^{d/\alpha} .
\end{equation} 
With \eqref{eq2} we obtain
\begin{equation}    \label{eq10} 
\comp^\ran (F_d^\alpha, \e) \le C \cdot \e^{-2} 
\end{equation}  
and if we only allow random bits then we obtain, using \eqref{eq3} 
and \eqref{eq8}, 
\begin{equation}   \label{eq11} 
\comp^{\rm coin} (F_d^\a , \e)\le C \, \frac{d}{\a} \, \e^{-2} \, (\log d + \log \e^{-1}). 
\end{equation} 

In the same way we obtain for the quantum computer the upper bounds
\begin{equation}    \label{eq12} 
\comp^\quant_\query (F_d^\alpha, \e) \le C \, \e^{-1}
\end{equation}  
and
\begin{equation}    \label{eq13} 
\comp^\quant  (F_d^\alpha, \e ) \le C \, \frac{d}{\a} \, \e^{-1} \, 
(\log d +  \log \e^{-1} ) . 
\end{equation}  

Observe that all these bounds \eqref{eq9}--\eqref{eq13}
are just upper bounds which we get by a particular proof technique. 
Actually it is known that the order in \eqref{eq9} is optimal, 
\begin{equation}    \label{eq14} 
\comp( F_d^\alpha , \e) \approx C_{d, \alpha} \,  \e^{-d/\alpha} ,
\end{equation} 
while the upper bounds for Monte Carlo methods are not optimal, we have 
\begin{equation}   \label{eq15} 
\comp^{\rm random}  (F_d^\a , \e) \approx C_{d, \alpha } \, \e^{-2d/(2\alpha +d)} 
\end{equation} 
and 
\begin{equation}   \label{eq16} 
\comp^{\rm coin}  (F_d^\a , \e) \le C_{d,\a}\, \e^{-2d/(2\alpha +d)}\, 
\log \e^{-1}.
\end{equation} 
For the proof of \eqref{eq15}
see Heinrich (1993), Novak (1988), or Traub, Wasilkowski, Wo\'zniakowski (1988). 
It is not difficult to show that \eqref{eq16} follows from \eqref{eq15}. 
Actually we will improve the exponent in the log-term slightly and prove such an 
upper bound with the factor $(\log \e^{-1})^{1/(1+2\a /d)}$, see \eqref{eq21}. 

\subsection{The problem and the results} 

Can the upper bounds \eqref{eq12} and \eqref{eq13} be improved, similarly  
as the upper bounds \eqref{eq10} and \eqref{eq11}? 
What is the optimal rate of convergence (or the rate of the complexity) for 
numerical integration with a quantum computer? 

In this paper we answer this question for classes such as the $F_d^\a$. 
We consider the more general H\"older classes 
$$
F_d^{k,\a} = \{ f: [0,1]^d \to \R \mid 
\Vert f \Vert_\infty \le 1, \ 
f \in C^k, \ 
|D^i f (x) - D^i f (y) | \le \Vert x - y  \Vert^\alpha , \ 
\forall \, D^i  \} , 
$$
where $D^i$ runs through the set of all partial derivatives of order $k$
and $k \in \N_0$, $0<\alpha \le 1$. For $k=0$ we obtain $F_d^{0,\a}=F_d^\a$. 
It is convenient to use the notation
$$
\gamma = \frac{k+\alpha}{d},
$$
because this number is a good measure for the smoothness and appears in all 
the estimates. 
First of all, the optimal orders for deterministic and (general) randomized methods 
are known, see, e.g., Novak (1988). We have
\begin{equation}    \label{eq17} 
\comp( F_d^{k, \alpha}  , \e) \asymp \e^{-1/\gamma}
\end{equation} 
and 
\begin{equation}   \label{eq18} 
\comp^{\rm random}  (F_d^{k, \a}  , \e) \asymp \e^{-2/(1+2\gamma)} .
\end{equation} 

Therefore we have to study only the quantities 
$\comp^\quant$, 
$\comp^\quant_\query$, and
$\comp^{\rm coin}$.
For the upper bounds we use a technique called ``variance reduction'' 
in the literature on Monte Carlo methods.
For the lower bound we use a decomposition technique of Bakhvalov, together 
with the lower bound of Nayak and Wu, see \eqref{eq4}. 
We obtain the following optimal rates of convergence. 

\medskip 

\begin{theorem} 
Define $\gamma= (k+\a)/d$, as above. 
\begin{equation}   \label{eq19} 
\comp^\quant_\query  (F_d^{k,\alpha}, \e) \asymp \e^{-1/(1+\gamma)}  , 
\end{equation} 
\begin{equation}   \label{eq20} 
\comp^\quant  (F_d^{k,\alpha}, \e) \le C_{d,k,\a} \, \e^{-1/(1+\gamma)} \, 
(\log \e^{-1} )^{1/(1+\gamma)} , 
\end{equation} 
\begin{equation}   \label{eq21} 
\comp^\coin   (F_d^{k,\alpha}, \e) \le C_{d,k,\a} \, \e^{-2/(1+ 2 \gamma)} \, 
(\log \e^{-1} )^{1/(1+ 2 \gamma)} . 
\end{equation} 
\end{theorem} 

\medskip 

To summarize the results one can say that 
\begin{itemize} 

\item
there is an exponential speed-up of quantum algorithms over 
deterministic (classical) algorithms, if $\gamma$ is small; 
the multiplicative speed-up is roughly $(1/\e)^{1/\gamma}$;  

\item
there is a (roughly) quadratic speed-up of quantum algorithms over 
randomized classical methods, if $\gamma$ is small.
\end{itemize} 

\subsection{Some comments} 

So far, most papers on quantum computing deal with discrete problems, such as 
factoring numbers or searching a database. 
Quantum computing also helps for the {\it continuous problems\/} of 
{\it numerical analysis\/} or {\it information-based complexity}. 
Grover (1998) studies, among other things, the computation of the 
mean of finitely many real numbers. 
Related problems and algorithms are investigated also in the papers 
Boyer, Brassard, H\o yer, Tapp (1998), 
Brassard, H\o yer, Tapp (1998), 
Grover (1996), and 
Mosca (1999).
The paper Nayak, Wu (1998) contains new lower bounds, while the recent 
paper
Brassard, H\o yer, Mosca, Tapp (2000)
contains new upper bounds.

Excellent surveys on quantum computing are Shor (1998) and 
Cleve, Ekert, Henderson, Macchiavello, Mosca (1999). 
Also the paper Abrams, Williams (1999) discusses 
the computation of sums and integrals, 
``as long as the function is not pathological''. 

In numerical analysis and information-based complexity we usually assume the real number 
model with an oracle that gives function values, see Traub, Wasilkowski, Wo\'zniakowski 
(1988) and, more formally, Novak (1995). 
For an expository account of continuous complexity on a classical computer see Traub, 
Werschulz (1998).
Concerning the ``allowed randomness'' of the algorithms we may distinguish between three 
different cases. 
If a random number generator is available that can produce random 
$\omega \in [0,1]$ according to the Lebesgue measure then we obtain the well known result
\eqref{eq18}. Also the other extreme case, where no randomness is available, is well studied 
and we obtain the result \eqref{eq17}. 
Hence we only have to consider the case of 
{\it restricted Monte Carlo methods\/} where coin tossing is allowed 
(and has unit cost), but not general random number generators. 
This case somehow corresponds to quantum computation where such a randomness can 
be easily realized. 

In Section \ref{grover} we give a 
little tutorial on quantum computation and present the search algorithm of Grover. 
We explain the model of computation and the cost of a quantum 
computation. Our proofs are contained in Section~\ref{proof}. 
We add a section where we discuss the computation of arbitrary bounded 
random variables by a quantum computer with a random number generator. 
Here we use a rather unrealistic model of computation because we assume that 
there is a random number generator without cost.

\goodbreak 

\section{The Model of Computation and the Search Algorithm of Grover}  \label{grover} 

In this section we describe the model of computation and 
the search algorithm of Grover (1996), see also 
Boyer et al. (1998). This section 
does not contain new results. 

Let $H_1$ be a 2-dimensional Hilbert space over $\C$ and let 
$e_0$ and $e_1$ be two orthonormal vectors in $H_1$. The space $H_1$ represents a quantum bit, 
in the Dirac notation we have 
$$
e_0 = \lg 0 \rs \quad \hbox{and} \quad e_1 = \lg 1 \rs .
$$
For $m \in \N$ quantum bits we use the $2^m$-dimensional tensor product space
$$
H_m= H_1 \otimes \dots \otimes H_1
$$
with $m$ factors. An orthonormal basis is given by the $2^m$ vectors 
$$
b_\ell  = e_{i_1} \otimes \dots \otimes e_{i_m} , 
$$
where $i_j \in \{ 0, \, 1 \}$ and 
\begin{equation}   \label{id} 
\ell = \sum_{j =1}^m i_j \, 2^{m-j} , \qquad \ell =0, \dots , 2^m -1. 
\end{equation}   
There are $2^m$ different $b_\ell $ and this corresponds to the $2^m$ different possibilities 
of an information that is given by $m$ classical bits. 
The Dirac notation for $b_\ell $ is just $\lg \ell  \rs$, instead of 
$e_{i_1} \otimes e_{i_2}$ one finds $\lg i_1, i_2 \rs$ or also 
$\lg i_1 \rs \lg i_2 \rs$. 
The formally different objects $(i_1, \dots , i_m)$ and $\ell$ or $b_\ell$ are often 
identified and called ``classical state''. 

One more piece of Dirac-notation is often used: $\lg x \rs \ls y \rg$ is a mapping, defined by
$$
(\lg x \rs \ls y \rg) ( \lg z \rs ) := (y, z) \cdot \lg x \rs . 
$$
Here we write $(y,z) = \ls y \mid z \rs$ for the scalar product. 
Therefore the projection $P_x$ on a normed vector $x$ is written 
as $\lg x \rs \ls x \rg$. It is defined by 
$y \mapsto (x,y) \, x$. 

The Fourier series of $x \in H_m$ is given by
\begin{equation}     \label{form} 
x = \sum_{i_j \in \{ 0,1\} } \a_{(i_1, \dots , i_m)} e_{i_1} \otimes \dots \otimes e_{i_m} =
\sum_{\ell=0}^{2^m-1} \b_\ell \, b_\ell .
\end{equation} 
We are only interested in normed vectors, $\Vert x \Vert =1$. 
All such vectors are called (pure) ``quantum states''. 
For each quantum state there is a probability distribution on the classical states: 
the probability of $\ell$ is $|\beta_\ell|^2$. 

A quantum algorithm starts with a classical state $k  \in \{ 0, \dots , 2^m-1\}$ which 
we identify with  $b_k \in H_m$. Then a number of unitary 
transformations $U_1, \dots , U_r$ are applied, 
the result is the quantum state
$$
x_k   = U_r \dots U_1 (b_k) 
$$
and can be written in the form \eqref{form}. Allowed are only those unitary transformations  
that are ``efficient'' in the sense that at most two quantum bits are changed. This means that, 
for example, $U_i$ changes the first two bits and is of the form 
$$
U_i (v_1 \otimes \dots \otimes v_m) = 
\wt U_i (v_1 \otimes v_2 ) \otimes v_3 \dots \otimes v_m ,
$$
for some unitary $\wt U_i : \C^4 \to \C^4$. 
In the quantum bit number model, which we use for the numbers $\comp^\quant$, 
one such unitary operation has cost one. 
The output of the algorithm, given by a final measurement, 
is a classical state $\ell \in \{0, \dots , 2^m-1 \}$, or certain bits of $\ell$. 
The probability of $\ell$ is $|\beta_\ell|^2$, where $\beta_\ell$ is the respective Fourier 
coefficient of $x_k = U_r \dots U_1 (b_k)$. 
We say that a quantum algorithm computes a given function if the probability 
of a correct output is at least $3/4$. 

Now we describe the search problem and {\it quantum computations with an oracle}. 
Let $m \in \N$ and $X_m=\{ 0, 1, \dots , 2^m-1 \}$. 
Assume that $f : X_m \to \{ 0, 1\}$ is an arbitrary mapping which, of course, 
can be identified with a subset of $X_m$. 
We define a corresponding unitary mapping $S_f$ on $H_m$ by 
$$
S_f(b_\ell) = -b_\ell \quad \hbox{if} \ f(\ell) = 1 \quad \hbox{and} \quad
S_f(b_\ell) = b_\ell \quad \hbox{if} \ f(\ell) = 0.
$$
We also put $S_0=S_f$ for $f(\ell) = \delta_{0,\ell}$. 
A {\it black box\/} or {\it oracle\/} $Q_f$ for $f$ is defined on $H_{m+1}$ 
by
$$
Q_f(b_\ell \otimes e_i ) = b_\ell \otimes e_{i+f(\ell)} .
$$
Here the plus sign in $e_{i+f(\ell)}$ means addition modulo 2,
also called exclusive or.  Then one can easily show that 
$$
S_f (b_\ell) \otimes e_0 = Q_f  P  Q_f (b_\ell \otimes e_0),
$$
and therefore the oracle $Q_f$ can be used to compute function values of $S_f$. 
Here $P$ is defined by 
$$
P(b_\ell \otimes e_i) = (-1)^i \, b_\ell \otimes e_i . 
$$
One can even simulate $S_f$ with $Q_f$ using only one application of $Q_f$: 
simply apply $Q_f$ on $(b_\ell \otimes (e_0 - e_1))$. 

{\it A search problem\/} is defined as follows. 
Let $F_m$ be the set of all $f_\ell : X_m \to \{ 0, 1 \}$ with $f_\ell (j) = 1$ 
iff $j = \ell$. 
Of course we may identify the sets $F_m$ and $X_m$, and to each $\ell \in X_m$
or $f_\ell \in F_m$ there is exactly one $b_\ell$. 
The problem is to find $\ell$ if $f=f_\ell$ is given by the oracle $Q_f$. 

The {\it algorithm of Grover\/} works as follows. First we define the Walsh-Hadamard 
transform $W_1: H_1 \to H_1$ by
$$
W_1(e_i) = \frac{1}{\sqrt{2}} ( e_0 + (-1)^i \, e_1 )
$$
and $W_m = W_1 \otimes \dots \otimes W_1$. Now the algorithm is defined by
\begin{equation}    \label{iter} 
(- W_m S_0 W_m S_f )^k \,  (W_m (b_0)) ,
\end{equation} 
where $k$ has the order $2^{m/2}$. 
It is shown in 
Boyer, Brassard, H\o yer, Tapp (1998) that $k \approx \pi \, 2^{m/2-2}$ 
is a very good choice that leads to a high probability of success.

The cost, in the quantum bit number model, of 
every iteration in \eqref{iter} is about $m$, the total cost 
(to find the element $\ell$ with high probability) is 
about $m \cdot 2^{m/2}$ or, with $N=2^m$, about $ \sqrt{N} \, \log N $. 
Formally the algorithm is slightly different because $S_f$ is only given 
by the oracle $Q_f$, hence we work with $m+1$ instead of $m$ quantum bits. 
We assume that an application of $Q_f$ has unit cost. 

For the problem ``compute the mean of $x_1, \dots , x_n$'' we assume, when we 
consider the quantum bit number model and the numbers $\comp^\quant$, 
that there is a Boolean 
oracle $Q$ that gives the $j$th bit (digit) of $x_i$.
For the quantum query complexity we allow real valued oracles and only 
count the number of oracle calls. 
In the case $0 \le x_\ell \le 1$ a real number quantum oracle has the form 
$$
Q ( b_\ell \otimes e_0 ) = 
b_\ell \otimes \sqrt{x_\ell} \, e_0 + 
b_\ell \otimes \sqrt{1-x_\ell} \, e_1 . 
$$
For all classical algorithms we allow real-valued oracles for the $x_\ell$. 

\section{Proof of Theorem 1}  \label{proof} 

We fix a space $F_d^{k,\alpha}$. 
First we prove the upper bounds for quantum computers. We use an algorithm 
of the form 
\begin{equation}  \label{eq22} 
A(f) = I(P_nf) + \wt Q_N^d (f-P_nf) .
\end{equation} 
Here $I(P_nf)$ is the integral of a function $P_nf$ 
and $Q_N^d$ is the $d$-dimensional midpoint rule, as in Section~1.2, 
which we apply to $(f-P_nf)$. 
We now explain the operators $P_n$ and $\wt Q_N^d$.
By $P_n$ we mean a projection operator by interpolation,
one can use piecewise polynomials, which uses $n$ function values 
and gives an order
$$
\Vert f - P_nf \Vert_\infty \asymp n^{-\gamma}, 
$$
for $f \in F^{k,\a}_d$. 
It is well known that this is the optimal order of convergence, see
Novak (1988). 
One evaluation of $(f - P_n f)$ can be implemented at a constant cost 
(where the constant depends on $d$ and $k$, but not on $n$). 

By $\wt Q_N^d$ we mean that we do not really apply the midpoint rule 
$Q_N^d$. Instead we evaluate this midpoint rule by a quantum computer 
up to some error $\e_1 \cdot n^{-\gamma}$ 
with the cost
$$ 
\cost^\quant_\query (N, \e_1 )  \le C \cdot \e_1^{-1}
$$ 
or 
$$ 
\cost^\quant    (N, \e_1) \le C \cdot
\e_1^{-1} (  \log N +  \log \log  \e_1^{-1} ) ,   
$$ 
respectively. 
The error of this method is bounded by
\begin{equation}   \label{eq23}
e(f) \le C \cdot N^{-\beta} + n^{-\gamma} \cdot  \e_1  , \qquad f \in F^{k,\alpha}_d , 
\end{equation} 
which is the sum of the error by discretization 
(the integral being replaced by $Q_N^d$) and the error made by the approximate 
evaluation of $Q_N^d$. 
This error bound is valid for all $\beta \le \alpha / d$, if $k=0$, and 
$\beta =1/d$, if $k>0$. 
In addition we have to assume that $N$ is at least of the order $n$. 
To simplify the presentation we use a $\beta$ which is always smaller than 1. 
The complete cost of the method is bounded by
\begin{equation}    \label{eq24} 
\cost^\quant_\query (N, \e_1 )  \le C \cdot ( n+  \e_1^{-1}) 
\end{equation}  
or 
\begin{equation}    \label{eq25} 
\cost^\quant    (N, \e_1) \le C \cdot  n +  
C  \cdot  \e_1^{-1} \, (  \log N +  \log \log \e_1^{-1} )  ,   
\end{equation}  
respectively. 
In the query-complexity case we simply choose $n \approx \e_1^{-1}$ and 
$$
N^{-\beta} \approx n^{-\gamma} \cdot  \e_1
$$
and observe that we may apply \eqref{eq23}, because of $\beta <1$. 
We obtain a cost of the order $n$ and an error of the order 
$n^{-\gamma -1}$ and so obtain the upper bound in \eqref{eq19}. 

In the quantum bit number model we take 
$$
n \approx \e_1^{-1} \, \log \e_1^{-1} , 
$$
again with 
$$
N^{-\beta} \approx n^{-\gamma} \cdot  \e_1. 
$$
Then we get the cost bound
$$
C \,  \e_1^{-1} \, \log \e_1^{-1}
$$
and the error bound 
$$
e(f) \le C \,  \e_1^{\gamma +1} \, ( \log \e_1^{-1} ) ^{-\gamma} . 
$$
We obtain 
$$
e(f)^{-1/(\gamma +1)}  \ge C \,  \e_1^{-1} \, ( \log \e_1^{-1} )^{\gamma /(\gamma +1)} 
$$
and therefore 
$$ 
\comp^\quant  (F_d^{k,\alpha}, \e) \le C \, \e^{-1/(1+\gamma)} \, 
(\log \e^{-1} )^{1 - \gamma /(\gamma +1)} .
$$

Now we prove the lower bound in \eqref{eq19}. 
The space $F_d^{k, \alpha}$ contains $n \asymp \e_1^{-1/\gamma}$ functions 
$f_1, f_2, \dots , f_n$ with disjoint supports such that

\begin{itemize} 

\item
$\int_{[0,1]^d} f_i \, dx = \e_1^{1+1/\gamma}$ and 

\item
$\sum_{i=1}^n \lambda_i f_i \in F_d^{k,\alpha}$ if $|\lambda_i | \le 1$, 

\end{itemize} 
see Novak (1988, p. 35). 
Consider now the following problem. Compute the mean value of the 
integrals $\int_{[0,1]^d} \lambda_i f_i \, dx$, where $|\lambda_i | \le 1$, 
up to some error $\e_2$. 
We can apply the lower bound of Nayak, Wu (1999), see \eqref{eq4}, 
to obtain the lower bound 
$$ 
\cost \ge C \, \min \,  ( \e_1^{-1/\gamma},  \, 
\e_1^{1+1/\gamma}  \, \e_2 ^{-1} )  . 
$$ 
Of course we put $\e_2=\e_1^{1+2/\gamma}$ and obtain 
$$ 
\cost \ge C \, \e_1^{-1/\gamma}  . 
$$ 
What we estimated was the cost to compute the mean value. 
Since $\int_{[0,1]^d} \sum_{i=1}^n \l_i f_i \, dx$ 
is actually the sum, the error is to be multiplied by $n$, 
hence $\e = \e_2 \, n \approx \e_1^{1+1/\gamma}$. 
Since the cost to obtain this error is at least of the order $\e_1^{-1/\gamma}$ 
we obtain 
$$ 
\comp^\quant_\query  (F_d^{k,\alpha}, \e) \ge C \, \e^{-1/(1+\gamma)} .  
$$

We finally prove the upper bound for restricted Monte Carlo. 
Instead of the arbitrary random numbers of a (general) Monte Carlo method
we can only use random bits or coin 
tossing. 
We use a discretized version of a well known variance reduction technique 
and write the method in the form
\begin{equation}  \label{eq26} 
A(f) = I(P_nf) + \wt Q_N^d (f-P_nf) ,
\end{equation} 
and only the meaning of $\wt Q_N^d$ is different from \eqref{eq22}. 
By $\wt Q_N^d$ we mean that we do not really apply the midpoint rule 
$Q_N^d$. Instead we evaluate this midpoint rule by the classical Monte Carlo 
method, again 
up to some error $\e_1 \cdot n^{-\gamma}$ 
with the cost
$$ 
\cost^\coin (N, \e_1 )  
\le C \cdot
\e_1^{-2} \, \log N + n  . 
$$
The bound $\e_1^{-2}$ is the classical Monte Carlo bound, the factor $\log N$ 
comes in because we need (about) $\log N$ random bits to select one node from 
the possible $N$ nodes. 
The error of this method is bounded by
$$ 
e(f) \le C \cdot N^{-\beta} + n^{-\gamma} \cdot  \e_1  , \qquad f \in F^{k,\alpha}_d , 
$$
see \eqref{eq23}. 
Again we use a $\beta$ which is always smaller than 1. 
We put $N^{-\beta} \approx n^{-\gamma} \,  \e_1$ 
and
$n \approx \e_1^{-2} \, \log \e_1^{-1}$
and obtain 
$$ 
\cost^\coin (N, \e_1 )  
\le C \cdot
\e_1^{-2} \, \log \e_1^{-1} 
$$
and
$$
e(f) \le C \,  n^{-\gamma} \, \e_1 \le C \, \e_1^{1+2\gamma} \, (\log \e_1^{-1})^{-\gamma}
$$
and therefore \eqref{eq21}. 

\section{A Remark on Randomized Quantum Algorithms} \label{random} 

We present an algorithm to compute the expectation of arbitrary bounded
random variables up to some error $\e >0 $. 
A classical randomized method needs time $\e^{-2}$. 
The proposed algorithm uses the algorithm of Grover and a random generator. 
We assume that the random generator is for free and this is certainly not 
a realistic assumption.
Let $(X, B, m)$ be a probability space 
and let
$$
F = \{ f: X \to \R \mid f \hbox{ is measurable and }  \Vert f \Vert_\infty  \le 1 \} . 
$$
We want to compute the integral of a function (or the expectation of a random variable) 
$$
I(f) = \int_X f(x) \, dm(x), 
$$
for $f \in F$. 
With \eqref{eq5} 
one obtains the complexity bound
\begin{equation}  \label{e5} 
\comp_\quant^{\rm random}  (F, \e)  \le C  \, \e^{-1} \, \log \e^{-1} , 
\end{equation} 
with a constant $C$ that does not depend on the particular space 
$(X, B, m)$. 
If we use the quantum query model, together with a free random generator, 
then we get in the same way the upper bound $C \, \e^{-1}$.  
We prove \eqref{e5}. 

\medskip

To compute an approximation $A(f, \e)$ of $I(f)$ for $f \in F$ and $0< \e \le 1/2$
(with the understanding that \eqref{eq6} should be true) 
we proceed as follows: 

\begin{itemize} 

\item
First we randomly select $x_1, \dots , x_{n} \in X$ using the random generator, where 
\begin{equation}          \label{e2}
n = \lceil 72 \, \e^{-2} \rceil . 
\end{equation} 
If we put\footnote{Of course $Q_n$ is simply the classical 
Monte Carlo algorithm for the computation 
of the integral. Here we only {\it define\/} the 
points $x_1, \dots , x_{n}$, we do not compute 
$Q_n(f)$. This expression is only used for the analysis and the intuition.} 
$$  
Q_n(f) = \frac{1}{n} \sum_{i=1}^{n} f(x_i) 
$$
then it follows 
from Chebyshev's inequality that 
\begin{equation}    \label{e3}  
|I(f) - Q_n(f)| \le \frac{\e}{3}  \quad \hbox{with probability at least $7/8$}. 
\end{equation}   

\item
We assume that an oracle is available for the computation of $f(x_i)$, up to a maximal 
error of $\e /3$. On input $i$ the oracle gives $\tilde f(x_i)$ such that 
$$
|f(x_i) - \tilde f(x_i) | \le \frac{\e}{3} .
$$

\item
Quantum computation. 
Let 
$$
\wt Q_n(f) = \frac{1}{n} \sum_{i=1}^{n}  \tilde f(x_i) . 
$$
Then we know $|Q_n(f) - \wt Q_n(f) | \le \e/3$ and therefore 
$$
|I(f) - \wt Q_n(f) | \le \frac{2}{3} \e 
\quad \hbox{with probability at least $7/8$}. 
$$
Hence we compute an approximation $A(f, \e)$ of $\wt Q_n(f)$ such that 
\begin{equation}  \label{e4} 
|A(f, \e ) - \wt Q_n(f) | \le \frac{\e}{3} 
\quad \hbox{with probability at least $7/8$}
\end{equation} 
and get \eqref{eq6}. 
With \eqref{eq5} and (31) one obtains the bound \eqref{e5} for the complexity
of the problem. 

\end{itemize} 

\subsection*{Acknowledgments} 

I thank Peter H\o yer very much for his helpful comments.
Peter gave me the reference [4] and we had a very interesting 
discussion about different models of [quantum] computation
and about different upper and lower bounds for the computation 
of the mean of $n$ numbers. Peter also was so kind to give me 
his results that will appear in [10]. 

I also thank several other referees and editors for valuable remarks. 
Peter Hertling found some minor mistakes that are corrected in this version. 
This work was done during my time as a fellow at the university in Leuven. 
I thank the colleagues from the Department of Computer Science, 
in particular Ronald Cools, for the kind hospitality. 

\parindent=0pt 

\def\BAMS{Bull. Amer. Math. Soc.\ }
\def\BIT{BIT\ }
\def\CO{Computing}
\def\CA{Constr. Approx.}
\def\JAT{J. Approx. Th.}
\def\JC{J. Complexity\ } 
\def\JMA{SIAM J. Math. Anal.}
\def\JMAA{J. Math. Anal. Appl.}
\def\JMM{J. Math. Mech.}
\def\MC{Math. Comp.}
\def\NM{Numer. Math.}
\def\RMJ{Rocky Mt. J. Math.}
\def\SJNA{SIAM J. Numer. Anal.}
\def\SR{SIAM Rev.} 
\def\TAMS{Trans. Amer. Math. Soc.}
\def\TOMS{ACM Trans. Math. Software}
\def\USSR{USSR Comput. Maths. Math. Phys.\ }  

\section*{References} 

\small 

\refx
D. S. Abrams, C. P. Williams (1999):
Fast quantum algorithms for numerical integrals and stochastic 
processes.
LANL preprint quant-ph/9908083. 

\refx
M. Boyer, G. Brassard, P. H\o yer, A. Tapp (1998):
Tight bounds on quantum searching. 
Fortschritte der Physik {\bf 46}, 493--505. 
See also LANL preprint quant-ph/9605034. 

\refx
G. Brassard, P. H\o yer, A. Tapp (1998):
Quantum counting.  
Lect. Notes on Comp. Science {\bf 1443}, 820--831. 
See also LANL preprint quant-ph/9805082. 

\refx 
G. Brassard, P. H\o yer, M. Mosca, A. Tapp (2000):
Quantum amplitude amplification and estimation.
LANL preprint quant-ph/0005055. 

\refx
R. Cleve, A. Ekert, L. Henderson, C. Macchiavello, M. Mosca (1999):
On quantum algorithms.
LANL preprint quant-ph/9903061. 

\refx
L. Grover (1996):
A fast quantum mechanical algorithm for database search. 
Proc. 28 Annual ACM Symp. on the Theory of Computing, ACM Press New York, 212--219. 
See also LANL preprint quant-ph/9706033 and Physical Review Letters {\bf 79}, 325--328. 

\refx
L. Grover (1998):
A framework for fast quantum mechanical algorithms.
Proc. 30 Annual ACM Symp. on the Theory of Computing, ACM Press New York. 
See also LANL preprint quant-ph/9711043 and Physical Review Letters {\bf 80}, 4329--4332. 

\refx
S. Haber (1970):
Numerical evaluation of multiple integrals. 
SIAM Review {\bf 12}, 481--526. 

\refx
S. Heinrich (1993):
Random approximation in numerical analysis. 
In: Functional Analysis, K. D. Bierstedt et al. (eds.), 
Marcel Dekker, New York, 123--171. 

\refx 
P. H\o yer (2000): 
Quantum complexity of the mean problem. 
Paper in preparation. 

\refx
M. Mosca (1999):
Quantum Computer Algorithms. 
Thesis. University of Oxford. 

\refx
A. Nayak, F. Wu (1998):
The quantum query complexity of approximating the median and related statistics. 
STOC, May 1999, 384--393. 
See also LANL preprint quant-ph/9804066. 

\refx
E. Novak (1983):
Eingeschr\"ankte Monte Carlo-Verfahren zur numerischen Integration. 
In: Proc. of the 4th Pannonian Symp. on 
Math. Stat., Bad Tatzmannsdorf, Austria, 
W. Grossmann et al. (eds.), 269--282.  

\refx
E. Novak (1988):
Deterministic and Stochastic Error Bounds in Numerical Analysis. 
Lecture Notes in Mathematics {\bf 1349}, Springer. 

\refx
E. Novak (1995):
The real number model in numerical analysis. 
J. Complexity {\bf 11}, 57--73. 

\refx
P. W. Shor (1998):
Quantum computing. 
Documenta Mathematica, extra volume ICM 1998, I, 467--486. 

\refx
J. F. Traub, G. W. Wasilkowski, H. Wo\'zniakowski (1988):
Information-Based Complexity. Academic Press. 

\refx
J. F. Traub, A. G. Werschulz (1998):                        
Complexity and Information. Cambridge University Press. 

\refx
J. F. Traub, H. Wo\'zniakowski (1992):
The Monte Carlo algorithm with a pseudorandom generator. 
Math. Comp. {\bf 58}, 323--339. 

\end{document}